\documentstyle[11pt,newpasp,twoside,epsf]{article}
\begin{document}
\title{Single Dish Calibration Techniques at Radio Wavelengths}
\author{K. O'Neil}
\affil{NAIC/Arecibo Observatory, HC3 Box 53995, Arecibo, PR 00612; koneil@naic.edu}

\begin{abstract}
Calibrating telescope data is one of the most important issues an observer faces.
In this chapter we describe a number of the methods which are commonly used to
calibrate radio telescope data in the centimeter wavelength regime.
This includes a discussion of the various methods often used in 
determining the temperature and gain of a telescope, as well as some of the
more common difficulties which can be encountered.
\end{abstract}

\section{Introduction -- The Importance of Calibration}

As you likely know, every telescope is unique.
One result of the uniqueness of individual telescopes is the difficulty of directly comparing
measurements from one telescope with those from another.  That is, Telescope A may 
record 280 counts for the peak of a given spectral line, while Telescope B may record
only 100 counts.  This is further complicated by the 
fact that even measurements taken on a given telescope, at a given
frequency, can change over time.  These changes can be the
result of changes in, e.g. the telescope system temperature, 
the telescope response, and/or
the atmospheric conditions.  This means that even if you only
observe your object with Telescope A,
but measure it repeatedly over a year, you may discover that your object's peak count varies
from 280 one day to 250 or 300 on another day, and so forth.
You then need to understand whether the source emission is truly
varying with time or if the differences are due to changes within
the telescope and equipment.

In order to compare measurements between two telescopes, or even between
one telescope taken at different times, we need a universal measurement system.
That is, we need to be able to state that 280 counts from Telescope A is equivalent to
X counts from Telescope B or equivalent to Y counts from Telescope A at a
different epoch.  This is the process of data (or telescope) calibration, and
the rest of this chapter will be devoted to presenting the various methods available, both
observationally and theoretically, to calibrate data.

\section{A Brief Review}

At this point it is useful to go over the more important equations
and concepts that are used in the calibration process.  All of these topics
are also covered in other chapters, where more detailed discussions
can be found.

\subsection{The Rayleigh-Jeans Approximation} 
Recall that the Planck Law for blackbody radiation is
\begin{equation}\rm B\:=\:{{2h\nu^3}\over{c^2}}\:{{1}\over{e^{h\nu/kT} - 1}}\end{equation}
where B is the brightness (specific intensity) measured in W/m$^2$/Hz/rad$^{2}$
(or Jy/rad$^{2}$); h is Planck's constant
(6.63 $\times10^{-34}$ Js); $\nu$ is the frequency in Hz;
k is Boltzman's constant (1.38$\times10^{-23}$ J/K);
and T is the temperature, in K. 
In the centimeter wavelength regime, it is often true that h$\nu\:\ll$ kT.
In this case, you can use the Taylor
Series to expand $e^{h\nu/kT} - 1$.
This provides the Rayleigh-Jeans approximation for blackbody radiation at radio
wavelengths:
\begin{equation}\rm B\:=\:{{2kT\nu^2}\over{c^2}}\:=\:{{2kT}\over{\lambda^2}}\: .
\label{eq:two} \end{equation}

For a discrete radio source of temperature T and subtending a
solid angle $\Omega_S$  the source flux density (S) in the Rayleigh-Jeans limit is
obtained by integrating over the source solid angle:
\begin{equation}\rm S\:=\:{{2k}\over{\lambda^2}}\:
\int_{\Omega_S}{T(\theta,\phi) d\Omega}\: .\end{equation}	
If the brightness temperature of the source is uniform across $\Omega_S$, this reduces to
\begin{equation}\rm S\:=\:{{2kT}\over{\lambda^2}}\:\Omega_S \: .\end{equation}

\subsection{Antenna Temperature}
The antenna temperature (T$_A$) can be defined as the
temperature of the antenna radiation resistance.
That is, let the telescope observe a point source (i.e. 
a source which is considerably smaller than the beam size)
which has a flux density S.  Then, replace the
feed of the telescope with a matched resistor (or load).  If you now adjust
the temperature of the resistor until
the power received is the same as it was for the point source
(observed with the antenna or feed horn),
the antenna temperature is equal to the resistor temperature.
That is, the measured spectral power is simply w=kT$_A$.

If absorbing matter is present, or the source does not completely fill the
beam, the measured antenna temperature will be less
than the source temperature.  In this case you
will measure an antenna temperature T$_A$ and you have
\begin{equation}\rm w\:=\:kT_A\end{equation}
\begin{equation}\rm S\:=\:{{2kT_A}\over{\lambda^2}}
\Omega_A\:=\:{{2kT_A}\over{A_e}}\: \end{equation}
\begin{equation}\rm w\:=\:{1\over 2}{A_e}S\;.\label{eq:six}\end{equation}
Here, A$_e$ is the effective aperture of the antenna,
and $\Omega_A$ is the solid angle of the telescope power pattern.
Recall that the antenna theorem gives A$_e\:\Omega_A\:=\:\lambda^2$
or, if ohmic losses are considered, A$_e\:\Omega_A\:=\:\epsilon_r=;\lambda^2$,
where $\epsilon_r$ is the fractional power transmission of the antenna, typically
close to unity.
(See the the chapters by, i.e. Hagen or Goldsmith for further derivation
of this quantity.)

In practice, the antenna temperature is given by
\begin{equation}\rm T_A\:=\:{{A_e}\over{\lambda^2}}
\int{\int{T_{source}(\theta,\phi)P_n(\theta,\phi)d\Omega}}
\:=\:{{\epsilon_r}\over{\Omega_A}}\int{\int{T_{source}
(\theta,\phi)P_n(\theta,\phi)d\Omega}}\: . \end{equation}
Here $\Omega_A$ is again the antenna solid angle and P$_n (\theta,\phi)$
is the antenna power pattern (normalized to unity at maximum).
If the source is a true ``point source'', i.e. 
it is small compared to the beam size, $P_n(\theta,\phi)\:\sim$ 1 
over the source solid angle and
\begin{equation}\rm T_A\:\approx\:{{\epsilon_r}\over{\Omega_A}}
\int{\int_{\Omega_S}{T_{source}d\Omega}}\:=\:\epsilon_R\;{{\Omega_S}\over{\Omega_A}}
\:T_{avg}\end{equation}
where $\rm T_{avg}$ is the brightness temperature averaged over the source.
If, on the other hand, the source is large compared to the beam size
and has a constant brightness temperature T$\rm _{const}$, the antenna
temperature is 
\begin{equation}\rm T_A\:=\:{{\epsilon_rT_{const}}\over{\Omega_A}}
\int{\int_{source}{P_n(\theta,\phi)d\Omega}}\:
=\:{{T_{const}}\over{\Omega_A}}\:\Omega_b^\prime \epsilon_r\: .\end{equation}
Here, $\Omega_b^\prime$ is the solid angle subtended by both the main beam and the
side lobes falling on the source.  (Note that if the source just fills the
main beam, $\Omega_b^\prime$ = $\Omega_m^\prime$, the main beam solid angle.)
Finally, if the source fills the sky
($\Omega_S\:\gg$ beam size), T$\rm _A$ = $\rm \epsilon_r$ T$\rm _{const}$.

\subsection{Minimum Detectable Temperature and Flux Density}
The minimum detectable antenna temperature is set by the fluctuations in the
receiver output caused by the system noise.  As has been discussed in other chapters,
this noise is directly
proportional to the system temperature (T$_{sys}$).  T$_{sys}$ can be  
broken down into three parts for analysis -- the antenna contribution (T$_{A}$), the receiver 
contribution (T$_{R}$), and the power loss between the two. 
More specifically, the system temperature can be written as
\begin{equation}\rm T_{sys}\:=\:T_A\:+\;T_{LP}\left[ 1/\epsilon\:
-\:1\right]\:+\:\left(1/\epsilon\right)\:T_R\end{equation}
where T$_{LP}$ is the physical temperature of
the transmission line between the antenna and the 
receiver, and $\epsilon$ is the efficiency of the transmission (0 $\le\:\epsilon\:\le$ 1).
The sensitivity of a radio telescope is then equal to
the rms noise fluctuations of the system:
\begin{equation}\rm \Delta T_{rms}\:=\:{{K_S T_{sys}}
\over{\sqrt{\Delta\nu\;t\;n}}}.\end{equation}
Here, $K_S$ is the sensitivity constant of the telescope (dimensionless and of order unity),
$\Delta\nu$ is the pre-detection bandwidth (in Hz), t is the integration time for one record
(in s), and n is the number of records averaged (dimensionless).
The minimum detectable temperature is typically considered to be 3--5 times the
rms noise temperature (e.g. $\rm \Delta T_{min}\:=\:3 T_{rms}$).

The minimum detectable temperature can be converted to a minimum brightness or flux density by
applying the Rayleigh-Jeans approximation (Equations \ref{eq:two} and \ref{eq:six}):
\begin{equation}\rm \Delta B_{rms}\;=\;{{2k}\over \lambda^2}\;{{K_S T_{sys}}\over{\sqrt{\Delta\nu\;t\;n}}}\end{equation}
\begin{equation}\rm \Delta S_{rms}\;=\;{{2k}\over A_e}\;{{K_S T_{sys}}\over{\sqrt{\Delta\nu\;t\;n}}}\: .\end{equation}

\section{System Temperature}
System temperature, for the purpose of this chapter, will be defined by
breaking it into two parts:
\begin{equation}\rm T_{sys}(\alpha,\delta,az,za)\:=\:
T_{off}(\alpha,\delta,az,za)\:+\:T_{source}(\alpha,\delta,az,za)\end{equation}
where $\alpha,\delta,az,za$ are the source right ascension, source declination,
and telescope position in azimuth and zenith angle, respectively.
The ``off source'' temperature is simply the temperature measured if the telescope
were pointed at a nearby region of blank sky:
\begin{equation}\rm T_{off}(\alpha,\delta,az,za)\:=
\:T_{RX}\:+\:T_{gr}(za,az)\:+\: T_{atm}(za)\:+\:
T_{CMB}\:+\:T_{BG}(\alpha,\delta)\; .\end{equation}
T$\rm _{RX}$ is the receiver temperature; T$\rm _{gr}$ the temperature contributed from the ground;
T$\rm _{atm}$ the contribution from the atmosphere; $T\rm _{CMB}$ the contribution from the
cosmic microwave background; and $\rm T_{BG}$ the contribution from background and foreground
celestial sources, including Galactic emission.

Theoretically, the system temperature of a telescope changes with telescope elevation
due to atmospheric emission following:
\begin{equation}\rm T_{atm}\:=\:T\left( {1\;-\;e^{-\tau A}}\right)\; ,\end{equation}
where $\tau$ is the atmospheric opacity at zenith 
and A is related to the zenith angle at which
the telescope is pointing (A = sec(za)).
A well-built telescope with an unblocked (or partially blocked)
aperture and a raised, movable dish can 
come close to achieving this theoretical model (Figure 1).
In reality, though, a number of factors must be added to this theoretical model.
In particular, most telescopes experience temperature changes due to changes in the 
ground radiation with elevation, reflection of radiation (both atmospheric and celestial) off
the telescope structure, and changes in the system itself.  

\begin{figure}
%\plotone{oneil_fig1.ps}
\caption{System temperature for the Green Bank Telescope as measured at 2 GHz
on 21-22 March, 2001.  (From Ghigo, Maddalena, Balser, \& Langston 2001.)}
%Figure 1
\end{figure}

In practice, the source temperature can be separated from the system temperature by
making two observations -- one which includes the source in the beam (the ON), 
and one which does not (the OFF).  The source temperature is then just:
\begin{equation}\rm T_{Source}\:=\:{T_{ON}\;-\;T_{OFF}}\end{equation}
\vskip 10pt
\begin{equation}\rm {T_{Source}\over T_{off}}\:=\:{{T_{ON}\;-\;T_{OFF}}\over{T_{OFF}}}\end{equation}
\vskip 10pt
\begin{equation}\rm T_{Source}\:=\:\left[ {{T_{ON}\;-
\;T_{OFF}}\over{T_{OFF}}}\right]\; T_{off} \: \label{eq:nineteen}\end{equation}
where $\rm T_{ON}$ and $\rm T_{OFF}$ are the power measurements for the two celestial positions.

At this point it should be clear that one of the most important measurements to be made
in order to understand, and calibrate, your data is the measurement of system temperature.
In the regime of centimeter wavelength astronomy there are two techniques commonly
used to make this determination.  These will be discussed in the next section.

\subsection{Switched Noise Diode}
The first technique for measuring system temperature 
uses a ``switched noise diode''.  With this
method, a noise diode with known effective temperature
at the desired frequency is 
coupled to the telescope system (Figure 2).  The telescope is then pointed to the blank sky
and two measurements are made -- one with the diode turned on ($\rm ON_{CAL}$) and one with the diode turned
off ($\rm OFF_{CAL}$).  These measurements are then used to 
determine the off-source system temperature, T$\rm _{off}$.   

\begin{figure}
\plotone{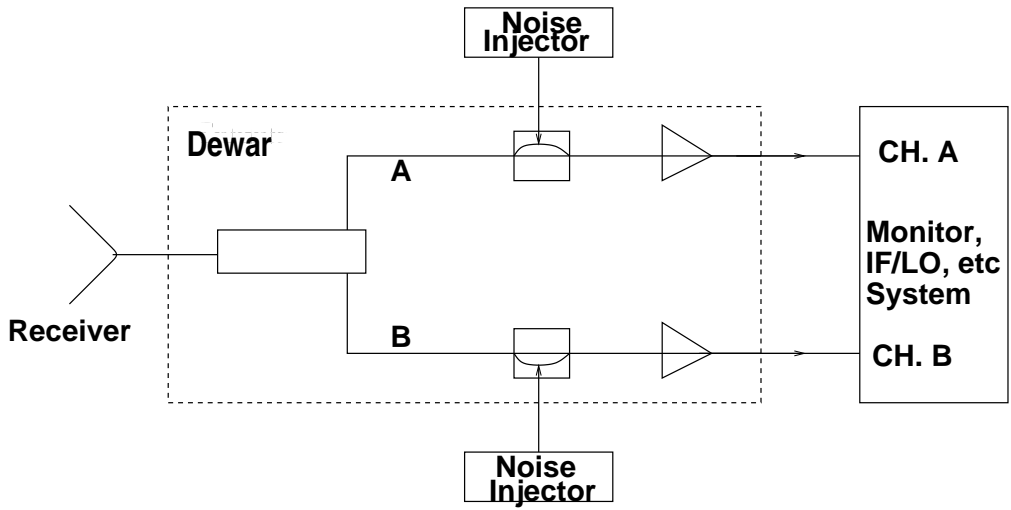}
\caption{Schematic drawing of the noise diode injection into
a receiver system.  This schematic is modeled after the L-Narrow 
receiver system at Arecibo Observatory.}
%Figure 2
\end{figure}

The basis behind the switched noise diode technique is quite simple.  The $\rm ON_{CAL}$
measurement is simply the sum of the system temperature plus the effective temperature of the
diode, while the $\rm OFF_{CAL}$ observation, in theory, measures only the system temperature.
The system temperature can then be derived by comparing the temperature
of the noise diode, measured as a fraction of the system temperature, with the known
temperature of the noise diode.  In other words, system temperature is
\begin{equation}\rm { {T_{off}} \over {T_{cal}}} \:=\:
{{OFF_{CAL}}\over{{ON_{CAL}}\;-\:OFF_{CAL}}}\end{equation}
\begin{equation}\rm T_{off}(K) \:=\: {{OFF_{CAL}}\over{ON_{CAL}\;-\:OFF_{CAL}}}
\:\times\: {\left[ {T_{cal}(K)}\; \right]}_{known}\end{equation}

Although in theory switched noise diodes provide an excellent tool for measuring
the system temperature, in practice a number of difficulties exist
which limit the accuracy of the noise diode method.  The first complication
is that the effective temperatures of most diodes have a frequency dependence.
This is not an overwhelming complication, as one can either 
obtain temperature measurements at a variety of frequencies  or use some
other multi-frequency calibration methods (i.e. van Zee, et al. 1997).  It is
merely an added issue to be aware of when
determining or applying noise diode measurements to your data.

A second issue to consider when using noise diodes to determine the system
temperature is the time stability of the diodes.   A good (well-built)
noise diode, in a well controlled environment, should remain stable for
a number of years.  Yet the performance of a diode should be checked by frequent (bi-annual
or more often) measurements of the diode response across the frequency band of
interest, to insure that the diode is performing according to its specifications.

The final issue to consider when using the noise diode system is the accuracy
of the measurement of the noise diode response.  This is a non-trivial issue,
as noise diode measurements are typically determined through bootstrapping off
another (`standard') noise diode, which is believed to have a well-known, 
accurately measured value.
Noise diode measurements therefore start off with an initial error from the
measurement of the standard diode.  As all other measurements are done through
bootstrapping off the first diode
measurement, this initial error is then propagated throughout all measurements.
Added to the initial measurement errors are, of course, the errors inherent in
the measurement process itself.  Finally, it should be remembered that 
the initial noise diode measurements (that is, the measurements coming from
the standard diode comparison) are not temperature measurements, but instead
most be converted from, i.e. a voltage measurement, 
and in that process yet more error is introduced.

In sum, the errors introduced just by the measurements of the noise diode 
value are:
\begin{equation}\rm \sigma_{total}\:=\:\sqrt{\sigma^2_{freq.\;dependence}\;
+\;\sigma^2_{stability}\;+\;\sigma^2_{measured\;value}
\;+\;\sigma^2_{conversion\;error}}\end{equation}
where
\begin{equation}\rm \sigma_{measure\;value}\:=\:
\sqrt{\sigma^2_{standard\;cal}\;+\;\sigma^2_{instrumental\;error}
\;+\;\sigma^2_{loss\;uncertainties}}\end{equation}
In spite of these difficulties, though, noise diodes can give a fairly
accurate indication of a telescope's system temperature, often to within 2\%
or less.   

{\bf Note --} System temperature is often measured at telescopes using a 
method known as the ``Y-Factor'' method.  This method employs two diodes (or similar sources)
of known effective temperatures and does not include any effects of the antenna.
If diode one has a known effective temperature T$_1$ and diode two
has a known effective temperature T$_2$, the ratio of the measured
power of the two diodes will be
\begin{equation}\rm Y\:=\:{{T_1\;+\;T_{off}}\over{T_2\;+\;T_{off}}}\end{equation}
where recall T$\rm _{off}$ for these purposes is the system temperature
without any effect of the antenna.  This ratio (the Y-Factor) can then be used 
to obtain T$\rm _{off}$ via
\begin{equation}\rm T_{off}\:=\:{{T_1\;-Y\;T_2}\over{Y\;-\;1}}\; . \end{equation}

\subsection{Hot \& Cold Loads}

Another method for obtaining the system temperature is through the use of hot and
cold loads.  A cold load is typically an absorbing system placed inside 
some gas or liquid at a known temperature (often liquid Nitrogen).  This 
load is then placed over, or otherwise coupled with,
the receiver of choice, and the power level
is measured.   The hot load is often a similar measurement of a load at ambient 
temperature.  Alternatively, the `cold' load can be a blank sky measurement and the
hot load can be a measurement of a load placed inside a liquid of known temperature.
These measurements can
then be used in the same fashion as a noise diode to obtain the system 
temperature.

Measuring the system temperature using hot and cold loads can be considerably more
reliable than the noise diode method.  The primary reasons for the superiority
of the hot/cold load method are that the temperature of the loads can be measured 
directly, and the measurements are all temperature measurements -- no
conversions are necessary.  However, the hot/cold load system is not
used at most centimeter wavelength radio telescope as building 
loads large enough to encompass a $\lambda\:>$ 5--6 cm feed,
and which have the easy availability necessary for on-the-fly measurements
is highly impractical.  As a result, the use of hot/cold loads to measure
system temperature on-the-fly is generally restricted to the $\lambda\:<$ 5--6 cm
regime.   In the $\lambda\:>$ 5--6 cm regime, hot/cold loads are often used
to check the noise diode measurements from the switched noise diode technique
(and can be used in place of a standard diode for calibration, as is the 
case at the Green Bank Telescope).
A more detailed discussion of this, and similar, calibration methods at short
wavelengths is given elsewhere in this book.

\section{Off-Source Observations}

In Section 3 we showed that the temperature of the object of interest 
can be found by measuring the
system temperature and using that measurement in combination with a blank-sky
measurement of T$_{sys}$, to calibrate the 
observations.  That is equation \ref{eq:nineteen} gave:
\begin{equation}\rm T_{source}(K)  \:=\: \left[{{T_{ON}\;-
\:T_{OFF}}\over{T_{OFF}}}\right]\;T_{off}(K)  \end{equation}
As Section 3 discussed the various means for determining the system temperature,
the next step is to determine the best method for obtaining an appropriate 
off-source, or blank sky, observation.

\subsection{Baseline Fitting}

Baseline fitting to just the  on-source records is 
potentially the simplest and most efficient of the methods for obtaining 
blank sky information for spectroscopic observations. 
The idea is straightforward -- a baseline is fit to
those spectral regions of an observation which are known to contain no signal
from the object of interest.  This baseline is then treated as the off source
(blank sky) observation (Figure 3a).

\begin{figure}
\centerline{
\plotfiddle{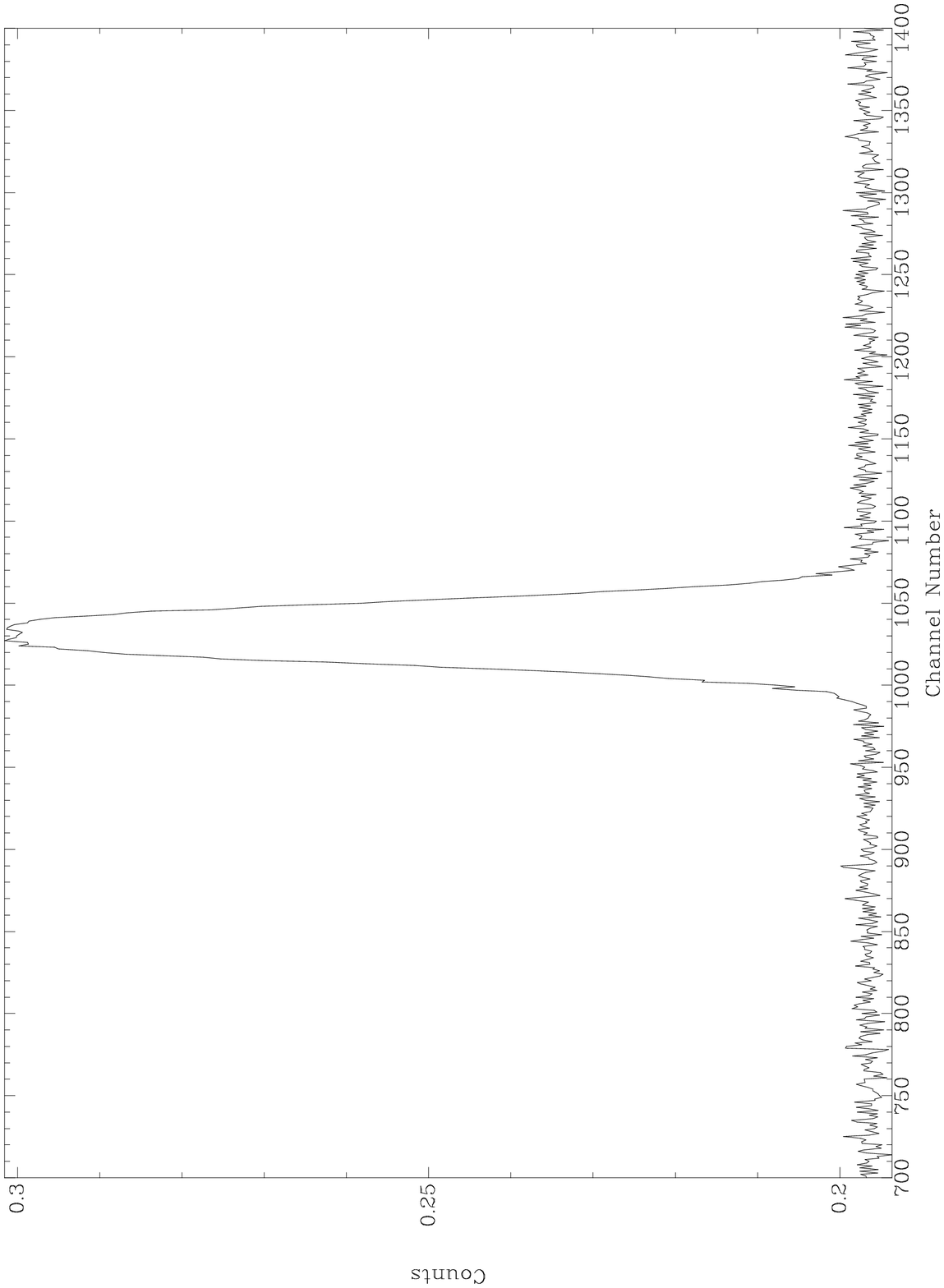}{2.00in}{-90}{25}{26}{-185}{160}
\plotfiddle{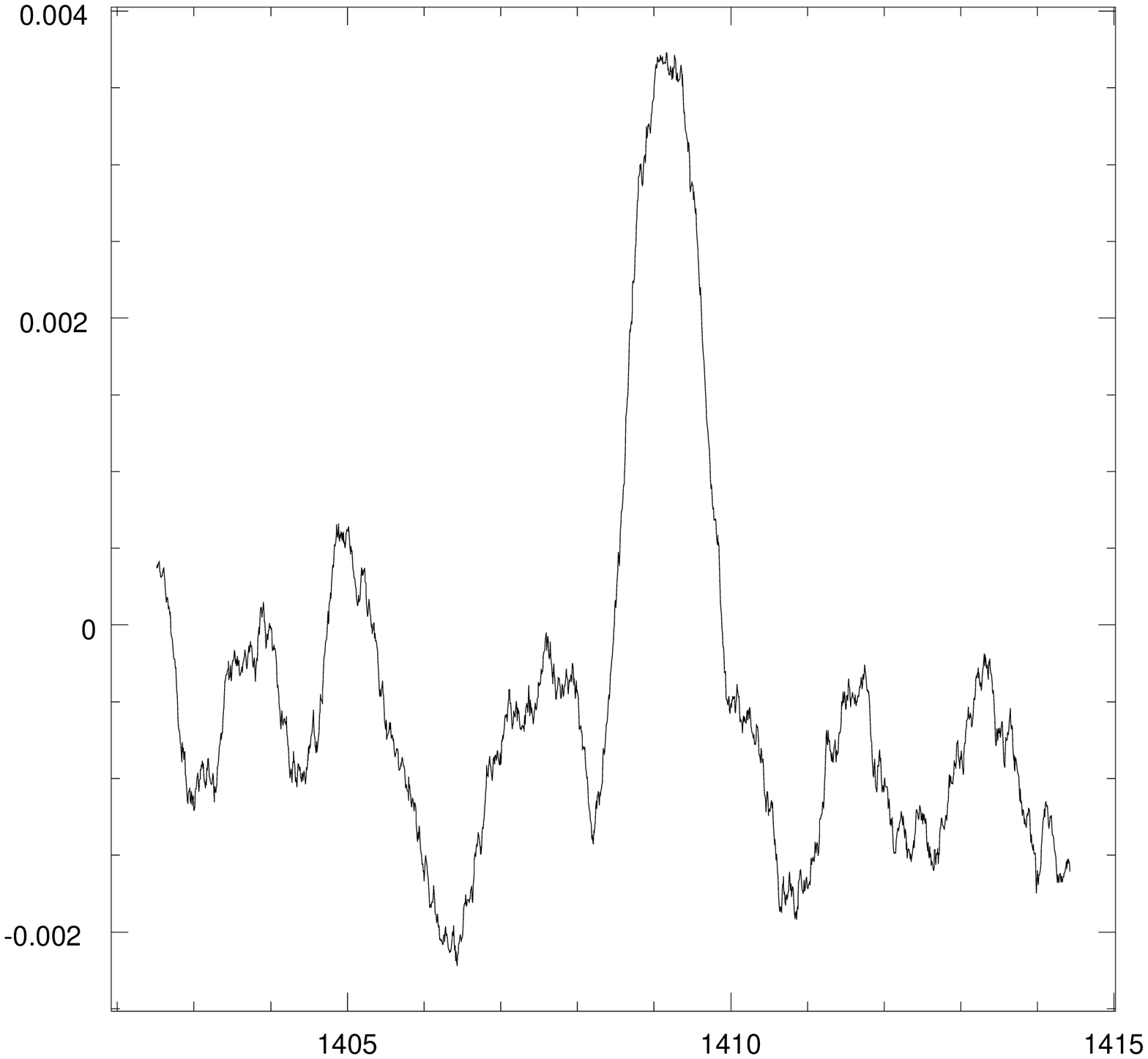}{2.00in}{0}{30}{27}{-375}{-40}
}
\caption{Examples showing successful fitting of a flat baseline to
an observation (left), and an observation in which baseline fitting is not
a viable option (right -- courtesy of C. Conselice). 
Both images include the spectral line of interest
and a number of `blank' channels.}
%Figure 3
\end{figure}

Although this procedure is extremely efficient, it has several drawbacks.
In particular, it is not a feasible option if the spectral line of interest 
is large compared to the bandpass (that is, if there are too few channels
from which to fit a reasonable baseline), or if there is are standing
waves present which have a frequency higher than, or of order of, the
bandpass of interest.  Additionally, if the baseline
is not reasonably flat for any reason this option should not be used, as a poor fit can
artificially add or subtract considerable emission to the line
of interest (Figure 3b). 

Errors induced from the baseline fitting
method come primarily from the quality of fit.

\subsection{Frequency Switching}

The frequency switching technique obtains blank sky information by
keeping the telescope pointed at the object of interest, but switching
the center frequency of the measurements (changing the frequency of the first
local oscillator).  As this mode of observing does not require any
movement of the telescope, it can be done very quickly and efficiently.
When the data from the two settings are subtracted, the quasi-stationary
effects introduced into the data after the first mixer, such as spectral
ripples, are cancelled.
Additionally, if the frequency is shifted such that the frequency range of interest
remains within the bandpass, but does not overlap its original 
``unswitched'' range, frequency switching can be
made extremely efficient (Figure 4).  Liszt (1997)
describes a deconvolution method for the case in which the two ranges
do overlap.

Frequency switching has a number of advantages over the 
baseline fitting method.  First, frequency switching reduces
the chance of error induced though poor fits to the baselines in the 
regions of interest.  Second, frequency switching can allow for higher
resolution spectra as considerable bandpass does not need to
be `wasted' to accommodate a large number of blank channels,
as is typically necessary for baseline fitting.
Finally, because frequency switching can occur at a very rapid rate
(on the order of a second, or less), frequency switching can 
cancel any post-mixer variations on this, or longer, time scales.

%\plotfiddle{oneil_fig4.ps}{2.0in}{-90}{40}{35}{-155}{190}
\begin{figure}
%\plotfiddle{oneil_fig4.ps}{2.0in}{0}{40}{35}{-115}{-60}
\caption{Data taken in frequency switching mode.  The center line
denotes the separation between the two images.  Here the x-axis is
channels and the y-axis in raw counts.  As the spectral line of interest
is available in both images, it would be possible to fold the 
data back together to achieve a higher signal to noise (and better
observing efficiency).}
%Figure 4
\end{figure}

The disadvantages to frequency switching at centimeter wavelengths
are few.  The primary difficulties are that (a) the 
redshift of the line of interest must be accurately known {\it a priori};
(b) the system must be stable enough that the baselines
of the primary observation and the frequency switched observation
are virtually identical; and (c) as with baseline fitting, if
there are significant standing waves in the baseline due to
reflections off the telescope structure for a partially blocked aperture
or the presence of strong continuum source within the beam, frequency 
switching may be unable to eliminate the standing pattern (provided,
of course, the characteristic frequency is not considerably wider than 
the frequency shift).

\subsection{Position Switching}

Position switching involves observing the object of interest for a fixed
period of time, and then moving the telescope to a blank sky region
to obtain the blank sky observations necessary for baseline subtraction.
Although costly in terms of telescope time, this observing method 
has the advantages that no {\it a priori} information (other than position
and a redshift known well enough that the line of interest will lie within
the spectrometer bandpass) need to
be known about the object and continuum emission within the beam can be
less of a problem (although see Section 4.4, below).

There are a number of issues to consider when deciding to use
position switching.  The first consideration is the time stability
of the telescope and baselines.  As position switching typically
requires re-pointing the telescope,
it is not feasible to switch between the on-source and blank-sky
observations at a rapid rate (although see the sections on wobbler
switching and chopper wheel techniques in the chapter by  Jewell).
This means the telescope baselines must remain stable over
a reasonable period of time, where this period of time
is defined by the amount of time it takes to observe both
the on and off-source positions.  (That is, if five minute
on- and off-source observations are being taken, with one
minute between the observations, 
the baselines must be stable for {\it at least} 11-12 minutes,
if not considerably longer.) 
%For this reason, standard
%position switching techniques are not feasible at short
%wavelengths where small changes in the atmosphere can 
%result in large changes in the baselines and system temperature,
%and the time scale for receiver stability is short.

A second issue which needs to be considered when performing
position switched observations is the dish illumination and
aperture blockage.
If the same portion of the dish is not illuminated 
during the on- and off-source observations, large differences
in the baselines can be present.  This is particularly true if there are
significant standing waves in the baseline due to a partially
blocked aperture.  These difficulties are
extreme for a telescope like the Arecibo 305-m
which has a fixed primarily reflector.  In this case, 
observations of different sky positions may not only illuminate
different areas of the reflector but can also have markedly 
different contributions from reflections off the ground and
telescope structure.  In this case it is 
prudent to obtain the off-source observation by
tracking a blank sky region chosen such that the illumination
pattern of the feed tracks across the same part of the
primary reflector as for the on-source observation.
In this case, if a five minute on-source observation of an object was taken,
starting when the object was at an Azimuth of 283\deg and
Zenith angle of 13\deg, the off-source
observation should also last for five minutes and start
at AZ , ZA = (283\deg, 13\deg).  Although this
method is time consuming, it can offer considerably
flatter the baselines and therefore 
a considerable reduction in the spectral deviations introduced 
by the system.

The final issue to consider with position switching
is that, if the size of the source is considerably more extended
in angular distance 
than the difference between the on and off-source sky positions (as is the case
for the ubiquitous Galactic HI), position switching becomes an impractical
option.  In these cases other alternatives such as frequency 
switching, baseline fitting, etc. must be considered.

\subsection{Variations}

To get around some of the difficulties inherent in
the above `standard' procedures for obtaining off-source
observations, a number of variations of these methods have
been (and are constantly being) devised.  
In this section we enumerate a few of these methods which have 
proven to be useful.

\subsubsection{Baseline Fitting with an Average Fit}

One method for reducing random noise which can cause difficulties when
fitting a function to the baselines (Section 4.1) is to average,
or median average, all the observations together and use that average to 
determine the baseline fit for the data.
This has the advantage of providing a fairly accurate baseline,
but at the expense of losing detailed information which
may be important in individual fits, particularly if the 
time over which the average is taken is longer than the stability 
time of the telescope system.

\subsubsection{Position Switching on an Extended Source}

As was mentioned in Section 4.3, the position switching method is 
not easily applied when observing a source which is extended when compared
with the offset distance between the on-source and off-source observations. 
Although frequency switching is typically the preferred observing method in these
cases, occasionally it is not a viable option.  In these cases, an
alternative observing method must be
considered.  One option which can be used when mapping 
an extended, but finite, source is to extend the map beyond the edges of the source.
The desired blank sky information is then obtained by averaging the 
off-source observations or by fitting  baselines across the map
using the off-source observations (Figure 5).

\begin{figure}
\plotfiddle{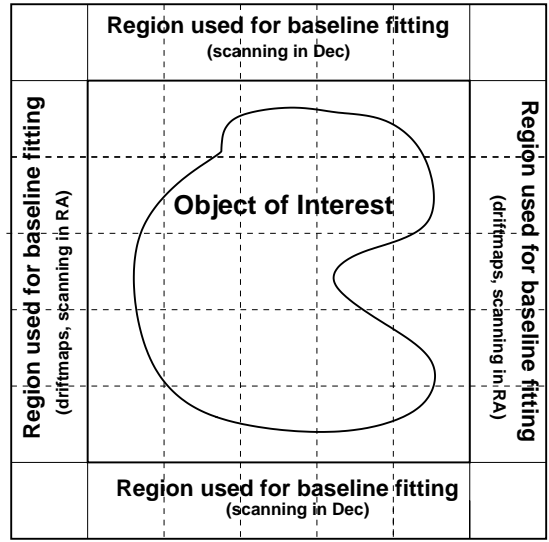}{3.00in}{0}{40}{40}{-100}{0}
\caption{Diagram showing the off-source regions typically used to
obtain blank sky information, using position switching, for
an extended source.}
%Figure 5
\end{figure}

If the telescope being used does not have a constant telescope illumination
as it points to different sky positions, such as is the 
case for the Arecibo 305-m telescope, it is often useful to 
use only those observations at a given Right Ascension (if the
map is stepping in Declination -- at a given Declination otherwise)
to obtain the best blank sky information. 
Even flatter baselines can be obtained in this case through
drift-scan mapping, as in this method the off- and on-source
observations are taken at the same (az,za).

\subsubsection{Position Switching in the Presence of Continuum Emission}

When position switching is attempted on a source with considerable
continuum emission (T$_{source}$ is a significant fraction of 
the value of T$_{sys}$),
a markedly different standing wave amplitude can be observed in the 
baselines between the on- and off-source observations.  As a
result, the components of the standing wave pattern produced by
the strong continuum emitter is not cancelled by the off-source
observation. This problem is particularly noticeable in telescopes
with partially blocked apertures and which have significant
standing waves even in the blank sky observations.

One method for dealing with this problem is to observe 
another continuum source, of similar strength to the object
of interest, and then to divide the source difference (ON $-$ OFF) spectrum by the
reference difference spectrum to eliminate the residual standing waves.
The result is a spectrum with a magnitude proportional to 
the ratio of the target and reference flux densities, and includes
any spectral-line component (emission or absorption) that may be present in the target:
\begin{equation}\rm R\:=\:{{\left({ON(\nu )\;-\;OFF(\nu )}\right)_{source 1}}\over
{\left({ON(\nu )\;-\;OFF(\nu )}\right)_{source 2}}}\end{equation}
In this case, the rms noise on the observed ratio is:
\begin{equation}\rm \sigma (R)\:=\:{{\sqrt{2}R}\over{\sqrt{\beta \tau}}}\:\times\;
\left[ \begin{array}{l}
{{SEFD^2_{S1,ON}\;+\;SEFD^2_{S1,OFF}}\over{(SEFD_{S1,ON}\;-\;SEFD_{S1,OFF})^2}}\\
\\ \;+\;
{{SEFD^2_{S2,ON}\;+\;SEFD^2_{S2,OFF}}\over{(SEFD_{S2,ON}\;-\;SEFD_{S2,OFF})^2}}
\end{array} \right] ^{1/2} \end{equation}
where S$_1$ and S$_2$ are continuum source one and two, respectively, and
SEFD is the system equivalent flux density of the telescope,
which is equal to the system temperature divided by the telescope gain
at the frequency of interest.
This method has been carefully tested  at Arecibo Observatory,
where the unique design of the 305-m telescope makes eliminating
standing waves from the baselines a challenge.  The results
show that, although this method of `double position switching' 
and obtaining the ratio of target flux densities 
initially increases the rms noise when compared to standard
position switched observations, the noise decreases as 1/$\sqrt{time}$
with continued observation, while the noise of the standard position switched
observations ceases to decrease after only four observation cycles
have been averaged (Figure 6). 
Further information on this method can be found in 
Salter \& Ghosh (2001), and references therein.

\begin{figure}
\plotfiddle{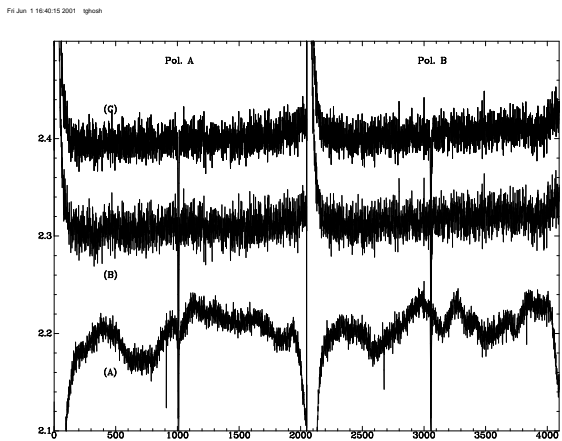}{2.0in}{-90}{30}{30}{-110}{160}
\caption{The average of three typical observing cycles with a strong continuum source;
(A) is for the standard position switching technique, (B) is double position switching
by dividing the [ON -- OFF]/OFF of the source with that of
the reference spectra, and (C) is double position switching using the 
double position switching method discussed in Section 4.4.  (Figure from
Salter \& Ghosh 2001.)}
%Figure 6
\end{figure}

\section{Flux Density Conversion}

At this point we have discussed both steps necessary for 
converting observed (raw) data into units of antenna temperature.
Although this is a significant step towards making data 
comparable between telescopes, it is not the complete picture.  
The reason for this is that every telescope has a different response,
or gain.  To complicate matters further, individual
telescopes also have different gains at different frequencies or even 
different elevation angles.
As a result, the temperature of a source must be converted
from observed antenna temperature to a gain-corrected measurement.
In radio astronomy this final, gain-corrected measurement is
typically in units of flux density per beam or main beam brightness.
Main beam brightness is simply the temperature measurement
corrected for telescope efficiency ($\rm T_{MB}\: = \:\eta_{beam} T_{measured}$).
As defined in Section 2, flux density per beam, on the other hand, is the integral of the 
source brightness over the telescope beam
\begin{equation}\rm S\:=\:{{2k}\over{\lambda^2}}\;
{{\int{\int{T(\theta,\phi)\;P_n(\theta,\phi)\;d\Omega}}}
\over{\int{\int{P_n(\theta,\phi)\;d\Omega}}}}\end{equation}
The unit of flux density is W m$^{-2}$ Hz$^{-1}$, or Janskies (Jy)
(1 Jy = 10$^{-22}$ W m$^{-2}$ Hz$^{-1}$).

As flux density per beam is the commonly used brightness
units for most centimeter wavelength
radio astronomy, the rest of the discussion in this section will concentrate
on calibrating data into flux density units.  If conversion into main beam
brightness is instead desired, the overall methods are the same.

\subsection{The Ideal Telescope}

In an ideal telescope -- one with a completely unblocked aperture, 
no ground reflection, lossless instrumentation, cables, etc., 
and a transparent atmosphere -- the telescope gain can readily
be theoretically modeled.  Even in the case of a fairly simple
and well understood system with minimum aperture blockage, an accurate
theoretical prediction of the telescope's gain can be obtained.
(See, e.g. the discussions by Lockman and Condon in this book).

As discussed by both Condon and Salter (this book),
if a telescope's response can be well modeled, the telescope
can be used to obtain absolute flux densities of continuum sources.
Then, if the absolute flux densities of a reasonable number of continuum
sources are determined at a variety of frequencies, 
monitored, and recorded, a catalog of `standard' continuum
sources can be developed.  These sources can then be used to
monitor telescope performance and look for any degradation in 
the telescope gain due to deterioration in the system components, distortion
of the reflector shape, etc.

\subsection{Bootstrapping}

For many telescopes, determining a telescope's gain theoretically is
extremely complicated due to considerable or irregular blockage of
the aperture, uncertain losses in the cabling or electronics,
uneven reflection from the ground, etc.  In these
cases, rather than relying upon what could be fairly inaccurate
models of the telescope gain, it is often good to
take advantage of pre-existing catalogs of standard continuum source
flux densities and ``bootstrap'' off those values to determine the gain
of the telescope of interest.

When choosing which sources to observe from source catalogs,
a few issues should be considered.  First, the size of the chosen source should
be small when compared with the size of telescope beam.  If the source
is of significant size when compared to the beam (a diameter more than $\sim$1/10
the telescope beam at the frequency of interest), issues such as the detailed
beam pattern and potential spillover of
the source onto the ground need be considered.  These problems can be eliminated
by insuring that the source is essentially point-like when compared to the beam.

The second item to consider when choosing a source for telescope
gain calibration is that the source should be strong enough to supply a fairly
high signal-to-noise ratio but not so strong as to cause
difficulties with baseline ripples or the telescope's dynamic range.

Finally, a good calibrator source should have a non-variable
flux density which has been well determined at the frequency of interest.
For a well defined telescope, errors in the previously determined
flux density measurements typically dominate all other errors inherent in determining
a telescope's gain using the bootstrap technique.  This means that
reducing the errors in the pre-determined flux density values results
in an essentially linear reduction in the errors in the telescope gain determination.
One of the results of this is that the error in determining of a telescope's gain
(when obtained through bootstrapping) can be greatly reduced 
by observing a number of sources at a range of positions in the sky.

The primary difficulty with using the bootstrap technique to 
determine telescope gain is that if a telescope's gain changes
significantly at different sky positions, or if the telescope's
beam illumination or aperture blockage changes as the telescope
points to different azimuth and/or zenith angles, it can 
readily become impractical for an observer to continuously 
monitor the telescope gain.  It may take more time to determine
the gain than to obtain the desired observations of the observer's
source.
In this case, the observer often needs to rely on a telescope's
pre-determined standard gain curves.

\subsection{Gain Curves}

When not available through theoretical models,
gain curves are obtained by observing a large number of standard 
flux density calibrators located across a telescope's visible sky
region at frequencies which range across the telescope's
usable bandwidth.  Gain  curves can be extremely useful
if a telescope's response is fairly stable over time and
the gain of a telescope varies across the sky.  

A telescope with an evenly illuminated
reflector typically has a gain curve which varies with 
telescope zenith angle in a fairly predictable manner (Figure 7).
Telescope gains can, though, become fairly complicated as
aperture blockage is increased, the illuminated 
portion of the main reflector changes (as is the
case with the Arecibo telescope), and the contribution from
ground reflection changes.  In this case it is possible to
obtain gain curves which vary as the telescope moves in azimuth as
well as zenith angle (Figure 8).

\begin{figure}
\plotone{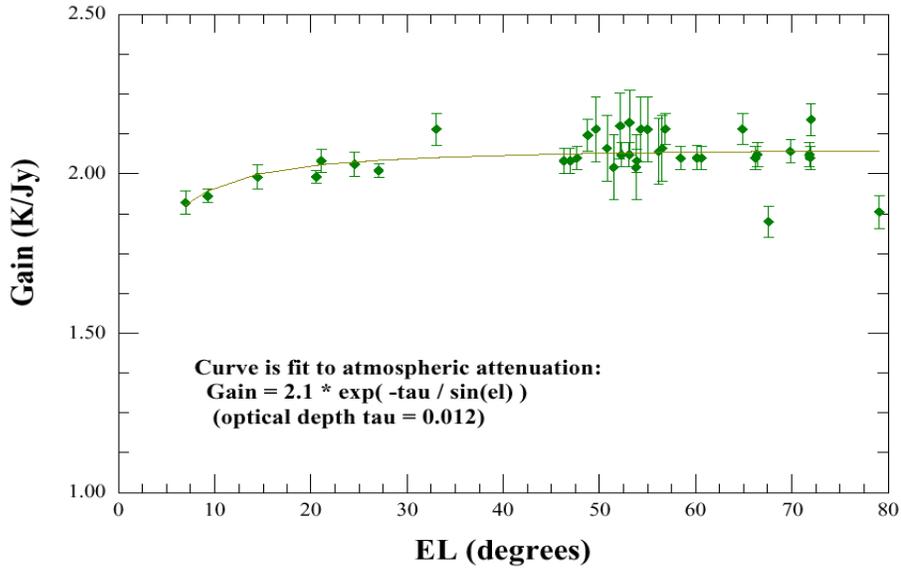}
\caption{Gain curve for the Green Bank Telescope.  (From Ghigo, et al. 2001.)}
%Figure 7
\end{figure}

\begin{figure}
\plotfiddle{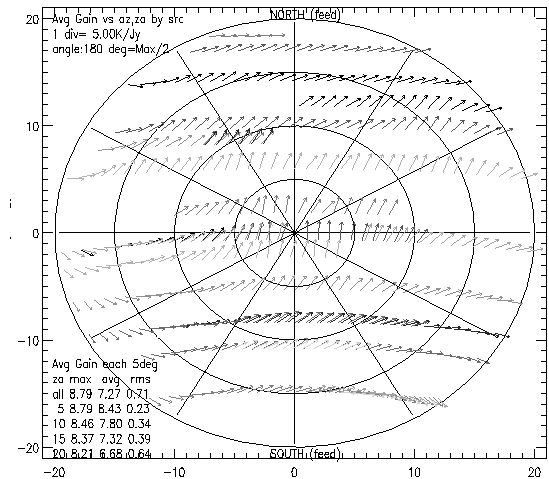}{2.2in}{0}{40}{40}{-110}{-80}
\caption{Gain distribution for the Arecibo 305-m telescope.  Here the both 
the length and angle of the arrow are proportional to the value of the gain,
and the polar plot is in terms of zenith angle (radius) and azimuth angle ($\theta$). }
%Figure 8
\end{figure}

Although the existence of pre-determined gain curves can
save the observer a considerable amount of telescope time, gain curves
should never be blindly accepted.  Instead, a circumspect observer
should recognize that although the shape of a telescope's gain curve
should remain relatively constant, a scaling of the curve may be necessary 
in the case of poor telescope focus, pointing offsets, or degradation
in the feed, electronics, cabling, etc.  As a result, it is always useful
to obtain observations of a number of flux-density standards 
at telescope positions near those of the objects of interest.
These observations can then be used to check the telescope gain
curve and, when necessary, to scale the curve accordingly.

\section{Other Issues}

There are a variety of other issues which can affect the gain and
system temperature of a radio telescope.  In this section we will describe
the more common issues an observer might encounter.

\subsection{Pointing and Focus}

Poor pointing or axial 
focus of a telescope can result in a reduction
of the telescope gain.  The reason for this is that a typical telescope
main beam pattern is Gaussian in nature (to a fairly low level). 
As a result, a relatively offset in pointing can result in a significant
reduction of the telescope gain
at the position of the object (Figure 9a).
In this case the signal-to-noise and calibration of the the observation
can suffer severely.  In a similar manner, poor telescope focus will
artificially diffuse an object so that less of the object falls onto the 
center of the beam where the telescope gain is at the highest (Figure 9b).  Again
this will result in a degradation in the object signal-to-noise ratio.

The method for checking axial focus is telescope dependent.
One reliable method for determining a telescope's pointing is by performing
cross-scans across a strong, point-like continuum source with an
accurately known position.  If a telescope's
pointing is accurate, the maxima of the scans will lie in the center of each
cross scan.  If the pointing is in error, an offset between the center of the
cross scans and the peak intensity measured by the observation will be seen.
In many cases, this offset can then be fed back into the telescope
pointing model.

\begin{figure}
\plottwo{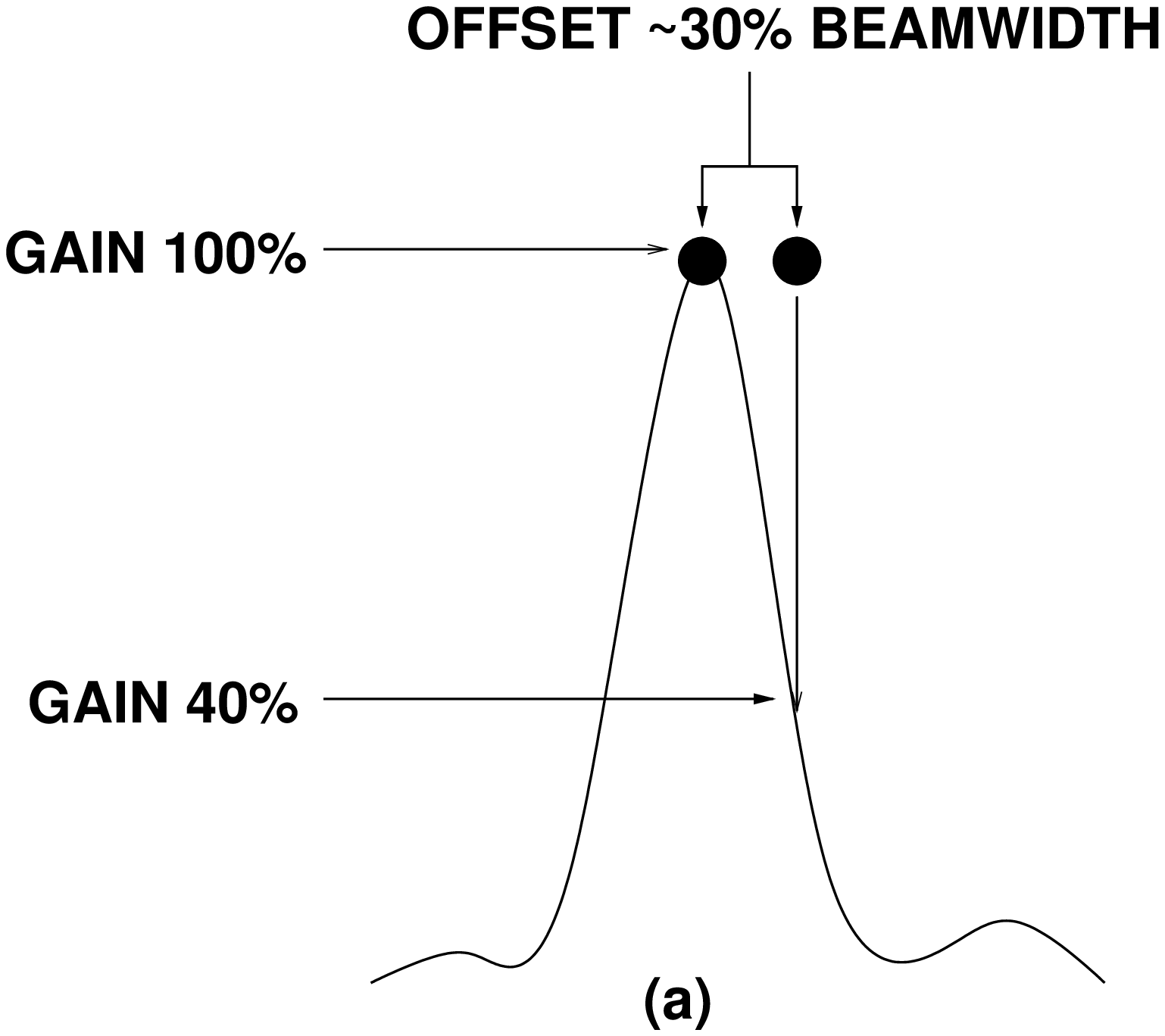}{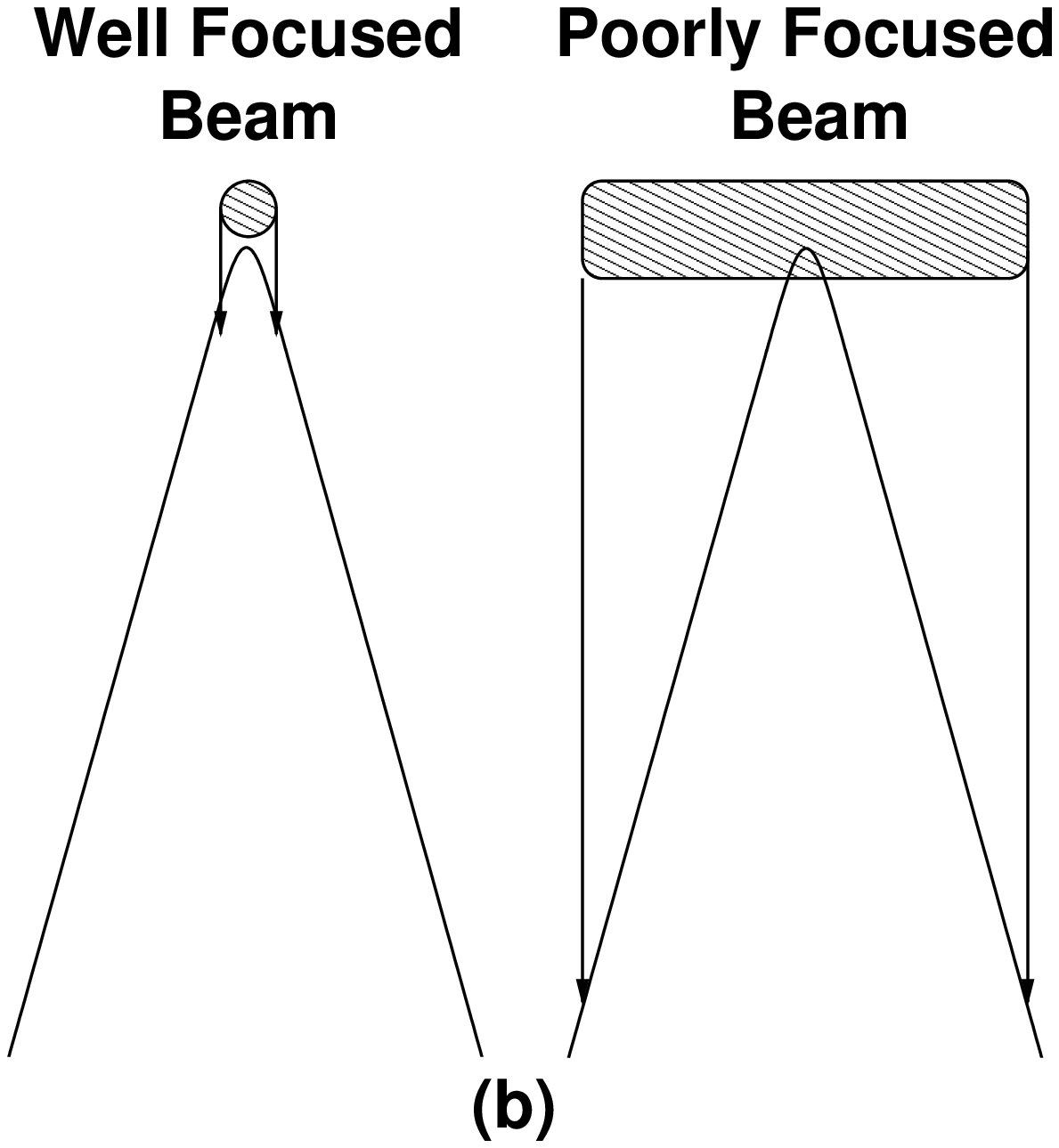}
\caption{Figure showing the results of (a) pointing errors and (b) focus errors
on the measured flux of an object.}
%Figure 9
\end{figure}

\subsection{Side Lobes}

Issues of side lobes in a telescope's beam pattern are
discussed extensively elsewhere in this book.  (See, e.g.,
the contribution by Lockman.)  However, as the presence of
side lobes can affect the calibration of an observation, it
is worthy of a brief mention here.
 
Because they can allow extraneous radiation to enter into
an observation, the presence of side lobes in a telescope's response
pattern can artificially increase the measured flux from a source.
This can result both in inaccurate measurements of a source's flux density
and, if a telescope's gain is determined through bootstrapping,
can create avoidable errors in the telescope's gain determination.
As a result, an observer should be aware of the presence and extent
of a telescope's side lobes at the frequency of interest.  If possible,
an observer should simply avoid observing any sky position
where the side lobe contributions will cause unacceptable errors
in the flux density measurements.  When this is not possible (much of the 
time, particularly when mapping an extended source),
the observer should attempt to use models of the side lobes and beam pattern to
deconvolve the side lobe contributions from the desired spectrum.

\subsection{Coma \& Astigmatism}

If a telescope's feed or secondary reflector is displaced or rotated from
its principal axis it will cause asymmetries in the detailed
telescope beam.  If a sub-reflector is shifted
perpendicular off the main reflector axis, a pointing error is
generated.  This is known as a comatic error.
Astigmatism occurs due to deformations in the reflector(s) and
results in irregularities (or lobes) on the beam pattern (Figure 10).
As with side lobes, the presence of
distortions in a telescope beam pattern can artificially increase the measured flux 
density of an extended source due to the addition of stray radiation,
and decrease the flux density for a point source
(due to decreased telescope efficiency).

\begin{figure}[h]
%\plotfiddle{oneil_fig10.ps}{2.8in}{0}{40}{35}{-120}{-20}
\caption{Beam response for the Arecibo 305-m telescope.  The contours are labeled in units
of percent peak intensity and are spaced by 0.4\%.
The dotted cross marks the position of intensity maximum.
The asymmetry in the beam pattern at the left 
of the image is due in large part to
the effects of the coma.  (From Heiles, et.al 2001).}
%Figure 10
\end{figure}

\section{Useful Resources}

In this section we list a few of the resources that are useful when calibrating
radio astronomy data.

\begin{itemize}
\item Baars, Genzel, Pauliny-Toth, \& Witzel, 1977 A\&A 61, 99\\[-10pt]
\item Baars 1973 IEEE Trans.  Trans. Antenna \& Propagation, AP-21, No. 4, pp. 461--474\\[-10pt]
\item Condon, et al. 1998, AJ, 115, 1693\\[-10pt]
\item Kraus, {\it Radio Astronomy} 1986 (Ohio:Cygnus-Quasar Books)\\[-10pt]
\item Kuhr, Witzel, Pauliny-Toth, \& Nauber, 1981 A\&AS 45, 367\\[-10pt]
\item Northern VLA Sky Survey, online at http://www.cv.nrao.edu/NVSS \\[-10pt]
\item Ott, et al. 1994 A\&A 284, 331\\[-10pt]
\item Tabara \& Inoue 1980 A\&AS 39, 379\\
\end{itemize}

\acknowledgements
Thanks to Chris Salter and Ron Maddalena for their helpful comments on this paper,
and of course many thanks to Paul M. for all his help.

\end{document}